\documentclass[letterpaper, 10 pt, conference]{ieeeconf}  

\IEEEoverridecommandlockouts                              

\usepackage{graphicx}
\usepackage{url}  

\usepackage{amsmath}

\title{\LARGE \bf
An Ensemble of Deep Learning Frameworks \\ Applied For Predicting Respiratory Anomalies
}

\author{Lam~Pham$^{1*}$, 
        Dat~Ngo$^{2*}$,
        Truong~Hoang$^{3}$, 
        Alexander~Schindler$^{1}$,       
        Ian~McLoughlin$^{4}$ %
\thanks{L. Pham and A. Schindler are with Center for Digital Safety \& Security, Austria Institute of Technology, Austria.}%
\thanks{D. Ngo is with School of Computer Science and Electronic Engineering, University of Essex, UK.}%
\thanks{T. Hoang is with FPT Software Company Limited, Vietnam.}%
\thanks{I. McLoughlin is with Singapore Institute of Technology, Singapore.}%
\thanks{(*) Main and equal contribution into the paper.}
}

\begin{document}

\maketitle
\thispagestyle{empty}
\pagestyle{empty}

\begin{abstract}

In this paper, we evaluate various deep learning frameworks for detecting respiratory anomalies from input audio recordings.
To this end, we firstly transform audio respiratory cycles collected from patients into spectrograms where both temporal and spectral features are presented, referred to as the front-end feature extraction.
We then feed the spectrograms into back-end deep learning networks for classifying these respiratory cycles into certain categories.
Finally, results from high-performed deep learning frameworks are fused to obtain the best score.
Our experiments on ICBHI benchmark dataset achieve the highest ICBHI score of 57.3 from a late fusion of inception based and transfer learning based deep learning frameworks, which outperforms the state-of-the-art systems.
\newline

\indent \textit{Clinical relevance}--- Respiratory disease, wheeze, crackle, inception, convolutional neural network, transfer learning. 
\end{abstract}
\section{INTRODUCTION}

Analysing respiratory sound has been recently attracted attention as leveraging robust machine learning and deep learning methods.
In these both approaches, systems proposed generally comprise two main steps, referred to as the front-end feature extraction and the back-end model.
In machine learning based systems, handcrafted features such as Mel-frequency cepstral coefficient (MFCC)~\cite{lung_tree_18, lung_hmm_02}, or a combined feature of four time domain features (variance, range, sum of simple moving average) and one frequency domain feature (spectrum mean)~\cite{lung_svm_01} are extracted at the front-end feature extraction.
These features are then fed into conventional machine learning models of Hidden Markov Model~\cite{lung_hmm_02}, Support Vector Machine~\cite{lung_svm_01}, or Decision Tree~\cite{lung_tree_18} for specific tasks of classification or regression.
Meanwhile, deep learning based systems make use of spectrogram where both temporal and spectral features are well represented.
These spectrograms are then explored by a wide range of convolutional deep neural networks (CNNs)~\cite{lung_cnn_01, lung_cnn_02,  pham2020_01, lam_11} or recurrent neural networks (RNNs)~\cite{lung_rnn_01, lung_rnn_02}.
Compare between two approaches, deep learning based systems are more effective and show potential results mentioned in~\cite{lung_cnn_01, pham2020_01, lam_11}.

As the deep learning approach proves powerful for analysing respiratory sounds, we evaluate a wide range of deep leaning frameworks which are applied for the specific task of audio respiratory cycles classification in this paper.
By conducting extensive experiments on the 2017 Internal Conference on Biomedical Health Informatics (ICBHI)~\cite{lung_dataset}, which is one of the largest benchmark respiratory sound dataset published, we mainly contribute: (1) We evaluate whether benchmark and complex deep neural network architectures (e.g. ResNet50, Xception, InceptionV3, etc.) are more effective than inception-based and low footprint models, and (2) we evaluate whether applying transfer learning techniques on the downstream task of respiratory cycle classification can achieve competitive performance compared with direct training approaches.

\section{ICBHI dataset and tasks defined}
ICBHI dataset~\cite{lung_dataset} comprising of  920 audio recordings was collected from 128 patients over 5.5 hours. 
Each audio recording contains one or different types of respiratory cycles (\textit{Crackle}, \textit{Wheeze}, \textit{Both Crackle \& Wheeze}, and \textit{Normal}) which are labelled with onset and offset time by experts. 
Given ICBHI dataset, this paper aims to classify four different types of respiratory cycles mentioned that is also the main task of the ICBHI challenge~\cite{lung_dataset}.
To evaluate, we follow the ICBHI challenge, then split 920 audio recordings into Train and Test subsets with a ratio of 60/40 without overlapping patient objects in both subsets (Note that some systems randomly separate ICBHI dataset into training and test subsets regardless patient object~\cite{lung_cnn_01, lung_cnn_02, lung_rnn_01}).
Using available onset and offset time, we then extract respiratory cycles from entire recordings, obtain four categories of respiratory cycles on each subset. 
Regarding the evaluating metrics, we use Sensitivity (Sen.), Specitivity (Spec.), and ICBHI scores (ICB.) which is an average score of Sen. and Spec..
These scores are also the ICBHI criteria as mentioned in the ICBHI challenge~\cite{icb_ratio} and~\cite{ic_cnn_19_iccas, pham2020cnnmoe}.

\section{Deep learning frameworks proposed}
\label{framework}

To classify four types of respiratory cycles from ICBHI dataset, we firstly propose a high-level architecture of three main deep learning frameworks as shown in Figure~\ref{fig:A1}: (I) The upper stream in Figure~\ref{fig:A1} shows how we directly train and evaluate small-footprint inception-based network architectures; (II) Benchmark and large footprint deep learning network architectures of VGG16, VGG19, MobileNetV1, MobileNetV2, ResNet50, DenseNet201, InceptionV3, Xception are directly trained and evaluated as shown in the middle stream of Figure~\ref{fig:A1}; and (III) The lower stream in Figure~\ref{fig:A1} shows how we reuse pre-trained models, which were trained with the large-scale AudioSet dataset, to extract embedding features for the training process on multilayer perception (MLP) network architecture.
In general, these three proposed deep learning frameworks comprise two main steps of the front-end spectrogram feature extraction and the back-end classification model.

\subsection{The front-end spectrogram feature extraction}
\label{feature}
\begin{figure}[t]
    	\vspace{-0.2cm}
    \centering
    \includegraphics[width =1.0\linewidth]{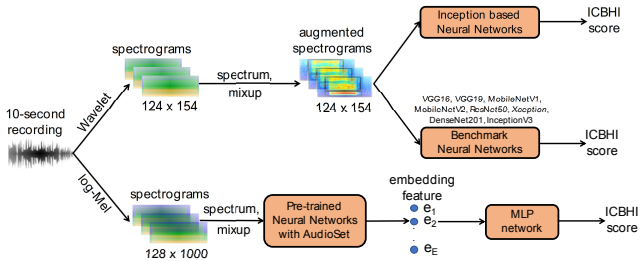}
    	\vspace{-0.5cm}
	\caption{The high-level architecture for three deep learning frameworks proposed.}
    \label{fig:A1}
\end{figure}
\begin{table}[t]
    \caption{The general inception based network architectures.} 
        	\vspace{-0.2cm}
    \centering
    \scalebox{0.95}{
    \begin{tabular}{|c |c |} 
        \hline 
            \textbf{Single Inception Layer}   &  \textbf{Double Inception Layers}  \\
        \hline 
             \multicolumn{2}{|c|}{BN} \\
        \hline 
             Inc(\textbf{ch1}) - ReLU  &    Inc(\textbf{ch1}) - ReLU - Inc(\textbf{ch1}) - ReLU \\
        \hline 
            \multicolumn{2}{|c|}{BN - MP - Dr(10\%) - BN} \\
        \hline 
             Inc(\textbf{ch2}) - ReLU    &   Inc(\textbf{ch2}) - ReLU - Inc(\textbf{\textbf{ch2}}) - ReLU\\
        \hline 
             \multicolumn{2}{|c|}{BN - MP - Dr(15\%) - BN} \\         
        \hline 
             Inc(\textbf{ch3}) - ReLU      &     Inc(\textbf{ch3}) - ReLU - Inc(\textbf{ch3}) - ReLU  \\
        \hline 
             \multicolumn{2}{|c|}{BN - MP - Dr(20\%) - BN}   \\         
        \hline 
             Inc(\textbf{ch4}) - ReLU & Inc(\textbf{ch4}) - ReLU - Inc(\textbf{ch4}) - ReLU\\
        \hline 
         \multicolumn{2}{|c|}{BN - GMP - Dr(25\%)}  \\              
        \hline       
         \multicolumn{2}{|c|}{FC(\textbf{fc1}) - ReLU - Dr(30\%)}       \\
        \hline          
         \multicolumn{2}{|c|}{FC(\textbf{fc2}) - ReLU - Dr(30\%)}       \\
        \hline          
         \multicolumn{2}{|c|}{FC(\textbf{4}) - Softmax}       \\
       \hline 
    \end{tabular}
    }
        \vspace{-0.3cm}
    \label{table:inc01} 
\end{table}
As proposed deep learning frameworks are shown in Figure~\ref{fig:A1}, we firstly duplicate the short-time cycles or cut off the long-time cycles to make all respiratory cycles equal to 10 seconds.
For the first two deep learning frameworks (I) and (II), we extract Wavelet-based spectrograms which proves effective in our previous work~\cite{lam_11}.
As we reuse the setting from~\cite{lam_11}, we then generate Wavelet spectrograms of $124{\times}154$ from 10-second respiratory cycles.
Meanwhile, we extract log-Mel spectrogram in the deep learning framework (III) as we re-use pre-trained models from~\cite{kong_pretrain} receiving log-Mel spectrogram input.
By using the same setting from~\cite{kong_pretrain}, we generate log-Mel spectrograms of $128{\times}1000$ for 10-second respiratory cycles.
To enforce back-end classifiers, two data augmentation methods of spectrum~\cite{spec_aug} and mixup~\cite{mixup1, mixup2} are applied on both log-Mel and Wavelet-based spectrograms before feeding into back-end deep learning models for classification. 

\subsection{The back-end deep learning networks for classification}

\begin{figure}[t]
    	\vspace{-0.2cm}
    \centering
    \includegraphics[width =0.9\linewidth]{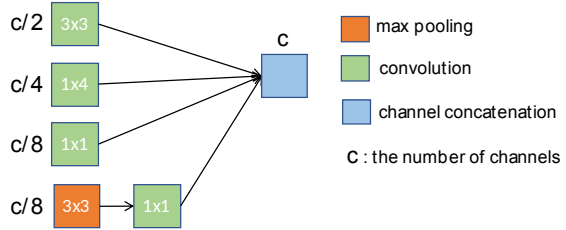}
	\caption{The single inception layer architecture.}
    \label{fig:A2}
\end{figure}

\textit{(I) The low-footprint inception based network architectures}: As the potential results were achieved from our previous work~\cite{lam_11}, we further evaluate different inception based network architectures in this paper.
In particular, two high-level architectures with single or double inception layers as shown in Table~\ref{table:inc01} are used in this paper.
These architectures perform different layers: inception layer (Inc(output channel number)) as shown in Figure~\ref{fig:A2}, batch normalization (BN)~\cite{batchnorm}, rectified linear units (ReLU)~\cite{relu}, max pooling (MP), global max pooling (GMP), dropout~\cite{dropout} (Dr(percentage)), fully connected (FC(output node number)) and Sigmoid layers. 
By using the two architectures and setting different parameters to channel numbers of inception layers and output node numbers of fully connected layers, we then create six inception based deep neural networks as shown in Table~\ref{table:inc02}, referred to as Inc-01, Inc-02, Inc-03, Inc-04, Inc-05, and Inc-06, respectively. 

\textit{(II) The benchmark and complex neural network architectures}: 
We next evaluate different benchmark neural network architectures of VGG16, VGG19, MobileNetV1, MobileNetV2, ResNet50, DenseNet201, InceptionV3, and Xception, which are available in Keras library~\cite{keras_app} and popularly applied in different research domains.
Compare with the inception based network architectures used in the framework (I), these benchmark neural networks show large footprint with complex architecture and trunks of convolutional layers.
\begin{table}[t]
    \caption{Setting for inception based network architectures.} 
        	\vspace{-0.2cm}
    \centering
    \scalebox{0.9}{
    \begin{tabular}{|c |c |c |c |c |c |c |} 
        \hline 
            \textbf{Networks}    &  \textbf{Inc-01}  &  \textbf{Inc-02} &  \textbf{Inc-03} &  \textbf{Inc-04} &  \textbf{Inc-05} &  \textbf{Inc-06}\\
                    \hline 
            \textbf{Single/Double}    &         &         &         &         &         &        \\
            \textbf{Inception }       & Single  & Double  & Single  & Double  & Single  & Double  \\
            \textbf{Layers}           &         &         &         &         &         &        \\
                    \hline 
            \textbf{ch1}              &  32     &  32     &  64     &  64     &  128    &  128 \\
            \textbf{ch2}              &  64     &  64     &  128    &  128    &  256    &  256 \\
            \textbf{ch3}              &  128    &  128    &  256    &  256    &  512    &  512 \\
            \textbf{ch4}              &  256    &  256    &  512    &  512    &  1024   &  1024 \\
            \textbf{fc1}              &  512    &  512    &  1024   &  1024   &  2048   &  2048 \\
            \textbf{fc2}              &  512    &  512    &  1024   &  1024   &  2048   &  2048 \\
       \hline 
    \end{tabular}
    }
    \vspace{-0.3cm}
    \label{table:inc02} 
\end{table}
\begin{table*}[t]
	\caption{Performance comparison of proposed deep learning frameworks.} 
        	\vspace{-0.2cm}
    \centering
    \scalebox{1.0}{
    \begin{tabular}{| l c | l c | l c | l c |} 
        \hline 
	    \textbf{Inception-based }  &\textbf{Scores}    &\textbf{Benchmark }   &\textbf{Scores}   &\textbf{Transfer learning}   &\textbf{Scores}   \\
  	    \textbf{Frameworks}  &\textbf{(Spec./Sen./ICB.)}    &\textbf{Frameworks}   &\textbf{(Spec./Sen./ICB.)}    &\textbf{Frameworks} &\textbf{(Spec./Sen./ICB.)}    \\

        \hline 
 Inc-01   &56.3/\textbf{40.5}/48.4   &VGG16        &70.1/28.6/49.3           &VGG14            &\textbf{82.1}/28.1/\textbf{55.1}\\     
 Inc-02   &69.7/31.9/50.8            &VGG19        &69.7/28.4/49.1           &DaiNet19         &76.4/26.9/51.7\\     
 Inc-03   &81.7/28.4/\textbf{55.1}   &MobileNetV1  &75.5/14.3/44.9           &MobileNetV1      &64.4/\textbf{40.3}/52.3\\     
 Inc-04   &\textbf{84.0}/24.8/54.4   &MobileNetV2  &74.7/16.1/45.4           &MobileNetV2      &76.0/32.7/54.4\\     
 Inc-05   &80.5/26.3/53.4            &ResNet50     &\textbf{88.0}/15.2/\textbf{51.6}  &LeeNet24         &70.7/30.9/52.8\\     
 Inc-06   &74.8/30.0/52.4            &DenseNet201  &71.7/30.3/51.1           &Res1DNet30       &74.9/26.7/50.8\\     
          &                          &InceptionV3  &70.9/\textbf{32.2/51.6}           &ResNet38         &71.6/32.2/51.9 \\     
          &                          &Xception     &75.7/22.1/48.9           &Wavegram-CNN     &69.0/38.1/53.5 \\

       \hline 
    \end{tabular}
    }
    \label{table:res_01} 
\end{table*}
\begin{table}[t]
    \caption{The MLP architecture used for training embedding features.} 
        	\vspace{-0.2cm}
    \centering
    \scalebox{1.0}{
    \begin{tabular}{l c} 
        \hline 
            \textbf{Setting layers}   &  \textbf{Output}  \\
        \hline 
         FC(4096) - ReLU - Dr($10\%$) &  $4096$       \\
         FC(4096) - ReLU - Dr($10\%$) &  $4096$       \\
         FC(1024) - ReLU - Dr($10\%$) &  $1024$       \\
         FC(4)    - Softmax & $4$ \\
       \hline 
    \end{tabular}
    }
    \vspace{-0.3cm}
    \label{table:mlp} 
\end{table}

\textit{(III) The transfer learning based network architectures}: As transfer learning techniques have proven effective for downstream tasks with a limitation of training data and smaller categories classified~\cite{kong_pretrain}, we leverage pre-trained networks which were trained with the large-scale AudioSet dataset from~\cite{kong_pretrain}: LeeNet24, DaiNet19, VGG14, MobileNetV1, MobileNetV2, Res1DNet30, ResNet38, Wavegram-CNN.
We then modify these networks to match the downstream task of classifying four respiratory cycles in ICBHI dataset.
In particular, we remain trainable parameters from the first layer to the global pooling layer of the pre-trained networks. 
We then replace layers after the global pooling layer by new fully connected layers to create a new network (i.e. Trainable parameters in new fully connected layers are initialized with mean and variance set to 0 and 0.1, respectively). 
In the other words, we use a multilayer perception (MLP) as shown in Table~\ref{table:mlp} (i.e. The multilayer perception (MLP) is configured by FC, ReLU, Dr, and Softmax layers) to train embedding features extracted from the pre-trained models (i.e. The embedding features are the feature map of the final global pooling layer in the pre-trained networks).

\section{Experiments and results}
\subsection{Experimental setting for back-end classifiers}
As using spectrum~\cite{spec_aug} and mixup~\cite{mixup1, mixup2} data augmentation methods, labels are not one-hot encoding format. Therefore, we use Kullback–Leibler divergence (KL) loss shown in Eq. (\ref{eq:kl_loss}) below.
\begin{align}
   \label{eq:kl_loss}
   Loss_{KL}(\Theta) = \sum_{n=1}^{N}\mathbf{y}_{n}\log \left\{ \frac{\mathbf{y}_{n}}{\mathbf{\hat{y}}_{n}} \right\}  +  \frac{\lambda}{2}||\Theta||_{2}^{2}
\end{align}
where  \(\Theta\) are trainable parameters, constant \(\lambda\) is set initially to $0.0001$, $N$ is batch size set to 100, $\mathbf{y_{i}}$ and $\mathbf{\hat{y}_{i}}$  denote expected and predicted results.
While we construct deep learning networks proposed in frameworks (I) and (II) with Tensorflow, we use Pytorch for extracting embedding features and training MLP in the framework (III) as the pre-trained networks were built in Pytorch environment.
We set epoch number=100 and using Adam method~\cite{Adam} for optimization.
%
\subsection{Performance comparison among deep learning frameworks proposed}

As the experimental results are shown in Table~\ref{table:res_01}, it can be seen that generally the low-footprint inception based frameworks and the transfer learning based frameworks are competitive and outperform the benchmark frameworks.
Table~\ref{table:res_01} records the best ICBHI score of 55.1\% from the Inc-03 framework and the transfer learning framework with the pre-trained VGG14 (i.e. The best performance obtained from the pre-trained VGG14 makes sense as this network outperforms the other network architectures for classifying sound events in AudioSet dataset). 
Notably, while we use the same network architecture of MobileNetV1 and MobinetV2, transfer learning frameworks significantly outperform the benchmark frameworks.
As a result, we can conclude that (1) applying the transfer learning technique on the downstream task of classifying respiratory cycles is effective; and (2) low-footprint inception based networks focusing on minimal variation of time and frequency are effective for respiratory sounds rather than benchmark and large network architectures.

\subsection{Early, middle, and late fusion of inception based and transfer learning based frameworks}
\begin{table}[t]
    \caption{Performance comparison of fusion strategies of inception based and transfer learning based frameworks.} 
        	\vspace{-0.2cm}
    \centering
    \scalebox{1.0}{
    \begin{tabular}{l c c c} 
        \hline 
            \textbf{Fusion Strategies}   &  \textbf{Spec.}  &  \textbf{Sen.} &  \textbf{ICB.}\\
        \hline 
         Pre-trained VGG14 &82.1 &28.1 &55.1\\
         Inc-03            &81.7 &28.4 &55.1 \\ 
         The early fusion  &79.9 &\textbf{30.9} &55.4 \\
         The middle fusion &\textbf{87.3} &25.1 &56.2 \\
         The late fusion   &85.6       &30.0     &\textbf{57.3}  \\
       \hline 
    \end{tabular}
    }
    \vspace{-0.5cm}
    \label{table:fusion} 
\end{table}
As the deep learning frameworks basing on Inc-03 and transfer learning with the pre-trained VGG14 achieve the best scores, we then evaluate whether a fusion of results from these frameworks can help to further improve the task performance.  
In particular, we propose three fusion strategies.
In the first and second fusion strategies referred to as the early and middle fusion, we concatenate the embedding feature extracted from the pre-trained VGG14 (e.g. the feature map of the global pooling of the pre-trained VGG14) with an embedding feature extracted from Inc-03 to generate a new combined feature. We than train the new combined feature with a MLP network architecture as shown in Table~\ref{table:mlp}.
While the feature map of the max global pooling (MGP) of Inc-03 is considered as the embedding feature in the first fusion strategy, the feature map of the second fully connected layer of Inc-03 (e.g. FC(fc2)) is used for the second fusion strategy.
In the third fusion strategy referred to as the late fusion, we use PROD fusion of the predicted probabilities obtained from these inception-based and transfer learning based frameworks.  
The PROD fusion result \(\mathbf{p_{f-prod}} = (\bar{p}_{1}, \bar{p}_{2}, ..., \bar{p}_{C}) \) is obtained by:
\begin{equation}
\label{eq:mix_up_x1}
\bar{p_{c}} = \frac{1}{S} \prod_{s=1}^{S} \bar{p}_{sc} ~~~  for  ~~ 1 \leq s \leq S,  
\end{equation}
where \(\mathbf{\bar{p_{s}}}= (\bar{p}_{s1}, \bar{p}_{s2}, ..., \bar{p}_{sC})\) is the predicted probability of a single framework, $C$ is the category number and the \(s^{th}\) out of \(S\) individual frameworks evaluated. 
The predicted label \(\hat{y}\) is determined by:
\begin{equation}
    \label{eq:label_determine}
    \hat{y} = arg max (\bar{p}_{1}, \bar{p}_{2}, ...,\bar{p}_{C} )
\end{equation}

As Table~\ref{table:fusion} shows, all three fusion strategies help to enhance the performance, report a further ICBHI score improvement of 0.3, 1.1, 2.2 from early, middle, late fusion strategies, respectively.
It proves that embedding features extracted from the inception based Inc-03 framework and the transfer learning framework with the pre-trained VGG14 contain distinct features of respiratory cycles.

\subsection{Performance comparison to the state of the art}
To compare with the state of the art, we only select published systems which follow the recommended official setting of ICBHI dataset~\cite{icb_ratio} with the train/set ratio of 60/40 and none of overlapping patient subject on both subsets.
As experimental results are shown in Table~\ref{table:res_02}, our system with a late fusion of inception-based and transfer learning frameworks outperform the state of the art, records the best score of 57.3\%.
\begin{table}[t]
    \caption{Comparison against the state-of-the-art systems.} 
        	\vspace{-0.2cm}
    \centering
    \scalebox{1.0}{

    \begin{tabular}{|l |l |c |c |c|} 
        \hline 
	    \textbf{Method}                    &\textbf{Spec.}   &\textbf{Sen.}   &\textbf{ICBHI Score}  \\
        \hline 
        HMM~\cite{ic_hmm_18_sp}                &38.0             &41.0           &39.0  \\              
        DT~\cite{lung_dataset}                 &75.0             &12.0           &43.0     \\        
        1D-CNN~\cite{t2021}                    &36.0             &51.0           &43.0  \\        
        SVM~\cite{ic_svm_18_sp}                &78.0             &20.0           &47.0     \\
        Autoencoder~\cite{dat_01}              &69.0             &30.0           &49.0     \\
        ResNet~\cite{ma2019}                   &69.2             &31.1           &50.2  \\ 
        ResNet~\cite{Ma2020}                   &63.2             &41.3           &52.3  \\ 
        Inception~\cite{lam_11}                &73.2             &32.2           &53.2  \\
        CNN-RNN~\cite{ic_cnn_19_iccas}         &81.0             &28.0           &54.0     \\ 
        ResNet50~\cite{microsoft2021}          &72.3             &40.1           &56.2  \\
        \hline 
        \textbf{Our best system}                    &\textbf{85.6}       &\textbf{30.0}     &\textbf{57.3}    \\
       \hline 
    \end{tabular}
    }
    \vspace{-0.3cm}
    \label{table:res_02} 
\end{table}
\section{Conclusion}
This paper has presented an exploration of various deep learning models for detecting respiratory anomaly from auditory recordings.
By conducting intensive experiments over the ICBHI dataset, our best model, which uses an ensemble of inception-based and transfer-learning-based deep learning frameworks, outperforms the state-of-the-art systems, then validate this application of deep learning for early detecting of respiratory anomaly.


\bibliographystyle{IEEEbib}
\bibliography{refs}
\end{document}